\documentclass[12pt,preprint,aps,showpacs,color,floatfix]{revtex4}
\usepackage{epsfig,colordvi}
\usepackage{epsfig,color}

\usepackage{epsfig}
\usepackage{graphicx}
\usepackage{graphics}
\newcommand{\etall}{$E_{\perp 12}$ }
\newcommand{\etqt}{$E_{\perp 12}^{QT}$}

\newcommand{\etps}{$E_{\perp 12}^{PS}$}
\newcommand{\et}{$E_{\perp 12}^{QT}$}

\newcommand{\be}{\begin{equation}}
\newcommand{\ee}{\end{equation}}
\newcommand{\bea}{\begin{eqnarray}}
\newcommand{\eea}{\end{eqnarray}}

\begin{document}
\title{Bimodality - a general feature of heavy ion reactions}
\author{A. Le F\`evre$^{1,2}$, J. Aichelin$^1$, C. Hartnack$^1$\\
and \\
 J.~{\L}ukasik $^{2,3}$, W.F.J.~M\"{u}ller $^2$, H.~Orth$^2$, C.~Schwarz$^2$, C.~Sfienti$^2$,
 W.~Trautmann$^2$, K.~Turz\'{o}$^2$, B.~Zwiegli\'{n}ski$^4$\\
A. Chbihi$^5$, J.D. Frankland$^5$, J.P. Wieleczko$^5$, M. Vigilante$^6$\\
((ALADIN and INDRA@GSI  Collaboration)}
\address{
$^1$ SUBATECH, Laboratoire de Physique Subatomique et des
Technologies Associ\'ees,
Universit\'e de Nantes - IN2P3/CNRS - EMN
4 rue Alfred Kastler, F-44072 Nantes, France.~ \\
$^2$ GSI Helmholtzzentrum f\"{u}r Schwerionenforschung GmbH,
D-64291 Darmstadt, Germany.~\\
$^3$ H. Niewodnicza\'{n}ski Institute of Nuclear Physics, Pl-31342 Krak\'{o}w, Poland.~\\
$^4$A.~So{\l}{}tan Institute for Nuclear Studies, Pl-00681 Warsaw, Poland.~\\
$^5$ GANIL, CEA, IN2P3-CNRS, B.P.~5027, F-14021 Caen Cedex, France.~\\
$^6$ Dipartimento di Scienze Fisiche, Univ.~di Napoli, I-180126 Napoli,
Italy.~\\}

\begin{abstract}
Recently, is has been observed that events with the {\it same} total
transverse energy of light charged particles (LCP) in the quasi
target region, \etqt, show two quite distinct reaction scenarios in the projectile
domain:
multifragmentation and residue production. This phenomenon has been
dubbed "bimodality". Using Quantum Molecular Dynamics  calculations
we demonstrate that this observation is very general. It appears in
collisions of all symmetric systems larger than Ca and at beam
energies between 50 A.MeV and 600 A.MeV and is due to large
fluctuations of the impact parameter for a given \etqt.
Investigating in detail the \etqt bin in which both scenarios are
present, we find that neither the average fragment momenta nor the
average transverse and longitudinal energies of fragments show the
behavior expected from a system in statistical equilibrium, in
experiment as well as in QMD simulations. On the contrary, the
experimental as well as the theoretical results point towards a fast
process. This observation questions the conjecture that the observed
bimodality is due to the coexistence of 2 phases at a given
temperature in finite systems.
\end{abstract}
\date{\today} \maketitle

\section{Introduction}
A while ago the INDRA and ALADIN collaboration has discovered \cite{tam} that,
in collisions of heavy ions -- Xe+Sn and Au+Au between $60$ and $100
A.MeV$ incident energy --, in a small interval of the total
transverse energy of light charged particles ($Z\le 2$) in the
quasi-target (QT) domain, \et , a quantity which is usually
considered as a good measure of the centrality of the reaction, two
distinct reaction scenarios exist. In this \et interval, in forward,
quasi-projectile direction, either a heavy residue is formed which
emits essentially light charged particles, or the system
fragments into several intermediate mass fragments. In the original
publication this phenomenon has been termed "bimodality" due to
reasons we will discuss further on. We stick to this name although our
interpretation of the origin of this phenomenon is different, and we
will call this interval in \et "bimodality interval". In the
meantime, this effect has also been observed by other groups
\cite{bru}.

This observation has created a lot of attention, because a couple of
years before theory has predicted \cite{gross,fg,cha} that in
finite size systems a first order phase transition weakens: in a
finite size canonical ensemble, which is determined by the
temperature T, the number of particles N and a given volume V (or a
given pressure p), it becomes more like a cross over. In infinite
matter the two phases coexist only at the transition temperature.
Below the transition temperature, $T_t (N,V)$, the system is in one
phase and above $T_t$ in the other phase. Because energy fluctuations are
suppressed by $\propto 1/\sqrt{N}$, this statement is also true if
the large system is treated micro-canonically. In finite systems,
the situation is different. In a canonical description of the
system, for a given T, N and V, the energy fluctuations can become
large, even larger than the finite latent heat. Therefore -- for a
given temperature close to $T_t$ -- for the same values of T, N and V, the
system can either be in the gas or in the liquid phase. This means
that if the system stays for long in thermal equilibrium, it moves
back and forth from one phase to the other. The simultaneous appearance of these
two modes (phases) for a given value of T,N,V has been called
"bimodality". In a micro-canonical description, bimodality is not
possible, because the energy of the two phases differs.

It is of course all but easy to identify these theoretical results
with observables obtained in a heavy ion reaction. Assuming that \et
is also a measure of the temperature of the system \cite{tam}, it is
nevertheless tempting to identify the residue with the liquid phase
of nuclear matter, and the creation of several medium or small size
fragments with the gas phase. The experimental observation of the
above mentioned bimodality scenario would then just be the
experimental confirmation of the theoretically predicted bimodality.

If this were true, the observation of bimodality would solve a
longstanding problem of heavy ion physics, the quest to identify the
reaction mechanism which leads to multifragmentation. This problem
arrived because many observables could be equally well described in
thermodynamical or statistical theories \cite{bon95,ber} as in
molecular dynamics type models \cite{aich,hartn,AMD,AMDind},
although the underlying reaction mechanism is quite different. The
statistical models assume that the system is in statistical
equilibrium when its density reaches a fraction of normal nuclear
matter density. Then, it suddenly freezes out and the fragment
distribution is determined by phase space at freeze-out. In
dynamical models, on the contrary, fragments are surviving initial
state correlations which have not been destroyed by hard
nucleon-nucleon collisions during the reaction, and equilibrium is
not established during the reaction. They can be already identified
very early in the reaction, when the density is still close to
nuclear matter density. A detailed discussion of how the reaction
proceeds in these models can be found in \cite{zbiri}.

Recently, it has been shown that the observation of bimodality alone
does not allow for the identification of the reaction mechanism.
Dynamical models describe the bimodality signal as well. In the
bimodality \et interval, they also show the presence of two
different event classes, and reproduce quantitatively the scaling
properties of \et \cite{arn}. Therefore, further studies of the
bimodality interval in \et are necessary to elucidate the reaction
mechanism.

The variables T, N and V determine the canonical ensemble
completely. If bimodality, in the sense of the coexistence of the
two phases at a given temperature, is at the origin of the
experimental observation, the values the average source velocity of
both modes as well as their temperature have to be
identical.

In this article, we analyze the experimental and theoretical events
which fall in the bimodality \et
interval, without further cuts, to study the average system
properties in this interval, and to investigate why bimodality,
defined as above as the observation of two different reaction
scenarios in a narrow interval of \et, is feasible even if the system
has not reached thermal equilibrium. In section II, we investigate in
detail the experimental observables in the bimodality \et bin and
whether they are compatible with the assumption that there are two
phases in thermal equilibrium. This detailed study is possible due
to the very good acceptance properties of the INDRA detector. In a
second step, we investigate in chapter III whether bimodality is a
phenomenon which occurs only in a small range of system sizes and
beam energies in which the system hits the transition temperature,
or whether it is a more general phenomenon. Especially, the energy
dependence is of interest, because with increasing energy there is a
change in the type of matter from which fragments are formed. At low
energies, it is the participant matter (the overlapping part of
projectile and target) which forms the fragments, whereas at
energies of a couple of hundreds of A.MeV, the fragments are formed
from spectator matter (the non overlapping part)\cite{zbiri,tsang}.

\section{system properties in the bimodality interval of \et}
For the investigation of the physics in the bimodality \et interval,
we follow the definitions of ref.~\cite{tam}. \et is defined as the
total transverse energy of particles with charge $Z\le 2$ on the
quasi target side $(\theta_{cm} \ge 90^\circ)$, calculated in the frame
in which the momentum tensor of all fragments with charge $Z \ge 3$
is diagonal. The diagonalization is done event by event. We define
$a_2$ as
\begin{equation}
a_2=(Z_{max1}-Z_{max2})/(Z_{max1}+Z_{max2})
\end{equation}
where $Z_{max1}$ is the charge of the largest fragment, while
$Z_{max2}$ is the charge of the second largest fragment, both
observed in the same event in the quasi-projectile (QP) hemisphere
-- at polar angles $\theta_{cm} < 90^\circ$ -- in the center of mass
of the system.
For a more accurate extraction of $a_2$, we reject events where less
than 70\% of the charge of the projectile has been detected.
Bimodality means that there exists a narrow interval in \et
in which events with large and small $a_2$ values are observed. In
this narrow transition region, we expect two types of events: One
with one big projectile residue accompanied with some very light
fragments (large $a_2$), the other with two or more similarly sized
fragments (small $a_2$). Events with intermediate values of $a_2$
should be rare.  In \cite{arn} we have studied $a_2$ as a function
of \et for the system Au+Au between 60 A.MeV and 150 A.MeV incident
energy that has been measured by the INDRA - ALADIN collaboration at
GSI \cite{luk}. We focus here on the data at 60 A.MeV and
concentrate on that experimental bimodality interval of \et where
the transition from small to large $a_2$ values occurs. We compare
the experimental events in this interval with filtered numerical
simulations for the same \et values. The filtering is done using a
software replica of the INDRA experimental set-up.

The simulations are performed with one of the dynamical models which
has frequently been used to interpret the multifragmentation
observables, the Quantum Molecular Dynamics (QMD) approach
\cite{aich,hartn,zbiri}. This approach simulates the entire heavy
ion reaction, from the initial approach of projectile and target up
to the final state, composed of fragments and single nucleons. Here,
nucleons interact by mutual density dependent two body interactions
and by collisions. The two body interaction is a parametrisation of
the Br\"uckner G-Matrix supplemented by an effective Coulomb
interaction. For this work, we have used a soft equation of state.
The initial positions and momenta of the nucleons are randomly
chosen and respect the measured rms radius of the nuclei as well as
the Fermi distribution in momentum space. Collisions take place if
two nucleons come closer than $r = \sqrt{\sigma / \pi}$, where
$\sigma$ is the energy dependent cross section for the corresponding
channel (pp or pn). The scattering angle is chosen randomly,
respecting the experimentally measured $d\sigma / d\Omega$.
Collisions may be Pauli blocked. For details we refer to
ref.\cite{aich,hartn}. For the later discussion, it is important to
note that, even for a given impact parameter, two simulations are
not identical, because the initial positions and momenta of the
nucleons as well as the scattering angles are randomly chosen. It
has been shown \cite{arn} that these simulations give a bimodality
signal in the same \et bin as the experiment and that the beam
energy dependence of this transition is reproduced.

\subsection{$a_2$ dependence of observables}
If the experimental signal, i.e. the existence of two distinct
reaction scenarios in a certain \et interval, reflects the
coexistence of two phases of a thermally equilibrated system, the
behavior of several observables can be predicted. In this case,
both phases  must have the same source velocity in the c.m.
and  the same temperature. Considering the nuclei resulting
from the decay of the projectile spectator as an ideal, classical,
non interacting gas, the mean kinetic energy of each fragment or
nucleon in the center-of-mass frame of this gas should be $3kT/2$.
This has also to be true for the largest fragment, called projectile
remnant (PR), in contradistinction to other theoretical approaches
in which the PR properties reflect the violence of the reaction,
which depends on centrality and hence on its size. The longitudinal velocity loss and the
transverse velocity of the heaviest fragment have been systematically studied for different
projectile-target combinations \cite{mor}.  The value of both
differs from the expectation for a heat bath particle, and depends
on the number of nucleons the projectile has lost in the course of
the interaction. This is one of the interests to study first the
properties of $Z_{max1}$ in the bimodality \et interval. The other is that
observables of the heaviest fragment should be give a clear signal of the
process because they are the least spoilt by eventual preequilibrium processes
which may disturb the light particle spectra.
\begin{figure}
\begin{center}
\includegraphics[width=8.6cm]{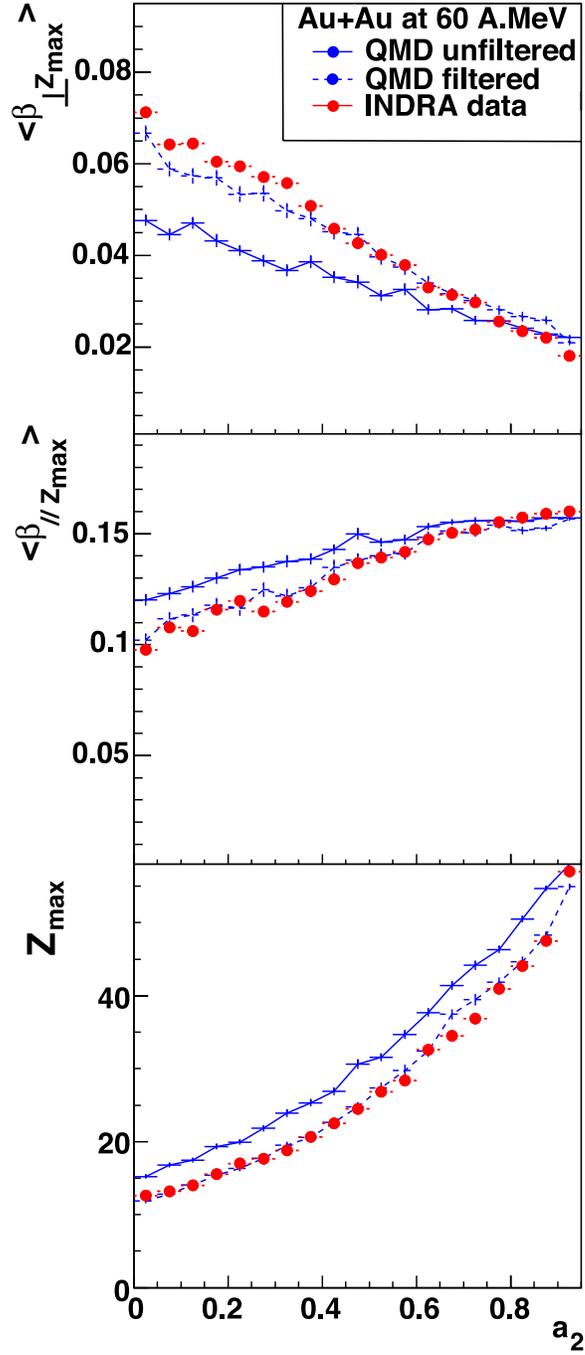}
\end{center}
\caption{(Color online) Average transverse velocity $\beta_\perp =
v_\perp/c$ (top) and $\beta_\parallel = v_\parallel/c$ (middle) of the QP fragment
($\theta_{cm} <90^\circ $) with the largest charge, $Z_{max1}$, as
well as $Z_{max1}$ (bottom) as a function of the QP $a_2$ for those events
which fall in the bimodality interval \etqt , see text. We
compare INDRA data (symbols) with filtered (dashed lines)
and unfiltered (full lines) QMD simulations for Au+Au collisions
at 60 A.MeV bombarding energy.}
\label{exZmax}
\end{figure}
Fig. \ref{exZmax} shows the transverse velocity $\beta_\perp =
v_\perp/c$ (top) and the longitudinal velocity, $\beta_\parallel =
v_\parallel/c$ (middle), in the c.m. of the system,
for the heaviest fragment in the QP region as a function $a_2$ and
for those fragments which belong to events which fall into the bimodality interval, $12 < E_{\perp\ 12}
^{QT}/E_{0cm}< 17$. $E_{0cm}$ is the incident energy per nucleon in
the c.m. system. The bottom panel displays the average
charge of the heaviest fragment as a function of $a_2$. In all
figures, we show the INDRA data as points and compare the
experimental results with unfiltered (full line) and filtered
(dashed line) QMD predictions. We see that the filter changes
substantially $\beta_\perp$ (due to a cutoff at small transverse
velocities) and to a less extend $\beta_\parallel$. After
filtering, the QMD calculations reproduce the trend of the data  but
overpredict the transparency, i.e. underpredicts $\beta_\perp$, for
small $a_2$ events corresponding to those with a small $Z_{max1}$.
If our assumption that \et characterizes the excitation energy and hence the
temperature of the system and that the events
in the bimodality bin have that temperature at which the system
can either in the gaz or in the liquid phase we expect that
$\beta_\perp$ as well as $\beta_\parallel$ -$\beta_{source}$
$\propto 1/\sqrt{mass}$. Here $\beta_{source}$ is the velocity of the
bimodal system in the c.m. which should be independent of $a_2$ because
the velocity of the bimodal system does not depend on which
mode is realized.

Consequently, this assumption does not offer
the possibility that  $\beta_\parallel$
increases with the particle mass or charge. The experimentally observed
linear increase of $\beta_\parallel$ with the mass of the fragment
follows, however, the trend already observed in central and
semi-central Xe+Sn events at 100 A.MeV \cite{frank} (where no
bimodality has been observed). The average velocity of the
heaviest fragment, in events with a smaller \et than at bimodality
follows this systematics as well. This linear decrease of the PR
velocity with decreasing mass is a very general phenomenon which has
first been studied by Morrissey \cite{mor} and complemented for higher beam energies by
Ricciardi et al. \cite{mosy}, although it has not been
shown yet that this systematics is still valid for such small PR
in this energy domain.
Such a linear dependence is expected if nucleons are removed
randomly from the cold projectile nucleus, under the condition that
they do not interact with the residue anymore. This removal leads to
a deceleration and an excitation of the remnant. Thus this increase
of $\beta_\parallel$ can be understood in models in which the
heaviest fragment is not in thermal equilibrium with the emitted
particles, but finds no explanation in purely thermal models.

In addition, we observe that the average transverse velocity
decreases with $a_2$. In order to discuss the compatibility with the
assumption of a canonical bimodal system, we have to consider the
unfiltered QMD events. They show an almost constant transverse
kinetic energy of 35 MeV, independent of the fragment mass and
charge, as it will be discussed at the end of section II.B. This
value confirms the analysis of \cite{wie} and is too large (even if
one considers a radial flow, as we discuss in the next section) to
be compatible with kT: the expected kinetic energy for fragments in
a thermal heat bath has to be smaller than the binding energy of
nucleons. Hence, in the bimodality \et interval, also $\beta_\perp$
is incompatible with the assumptions that two phases are present,
having both the same temperature.

Thus neither the mass dependence of $\beta_\parallel$ nor that of
$\beta_\parallel$ of  the heaviest fragment $Z_{max1}$ in the
bimodality \et interval are compatible with the expectations for
a finite system in which two phases are in equilibrium. On the
contrary, they follow the systematics which we have observed for
other \et regions where no experimental signs of the presence of two
phases in thermal equilibrium, i.e. bimodality behavior, are found.
These properties can be explained in models which are genuinely
non-equilibrium and which have successfully been applied to interpret
data in many experimental situations.

\subsection{Particle properties in the \et interval which shows bimodality}
\begin{figure}[ht]
\begin{center}
\includegraphics[width=16.8cm]{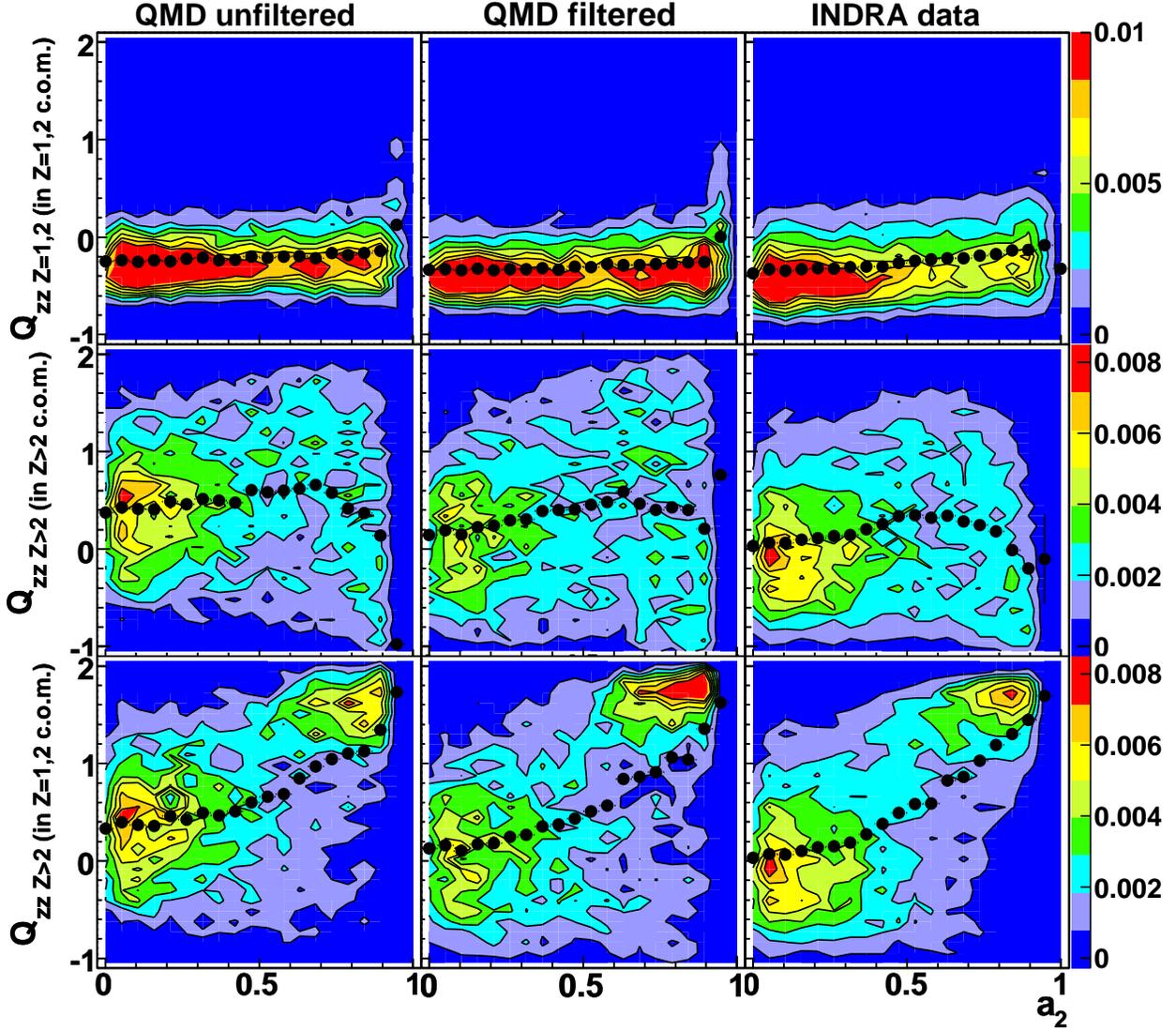}
\end{center}\emph{}
\caption{(Color online)Distribution of $Q_{zz}$ for Au+Au events at 60 A.MeV
which fall in the \et bimodality bin.
We display the $Q_{zz}$ distribution (see text)
for light charged particles (Z=1,2) (top row) in their own
center-of-mass,
and for fragments with charge $Z\ge 3$ in their own (middle
row) and in the  Z=1,2 c.m. system (bottom row),
as a function of $a_2$. The
points are the mean values of $Q_{zz}$ as a function of $a_2$.
>From left to right, we display unfiltered and filtered QMD
predictions, and the INDRA data, respectively.
The points mark the mean values.} \label{figqzz}
\end{figure}
\begin{figure}[ht]
\begin{center}
\includegraphics[width=8.6cm]{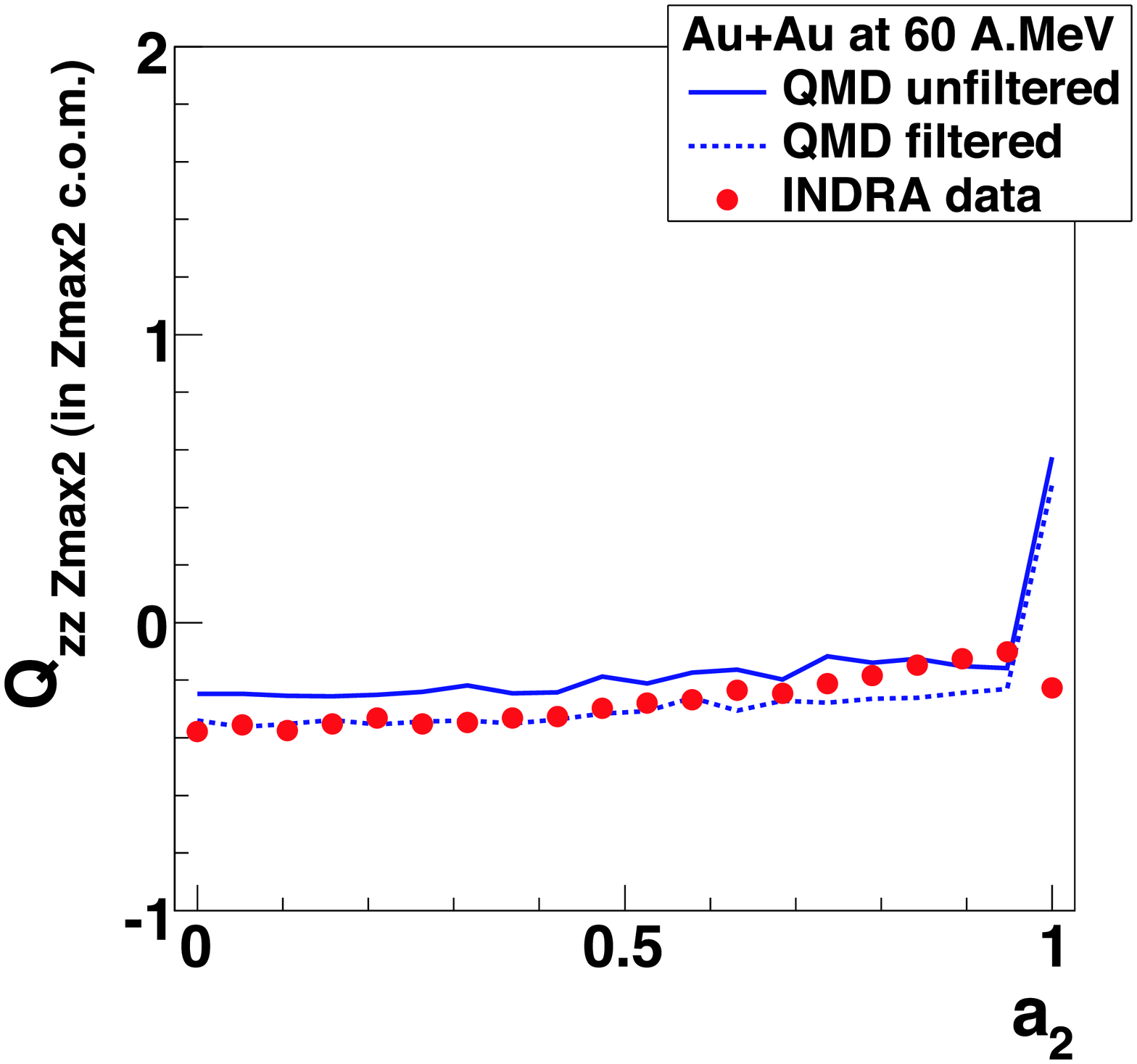}
\end{center}\emph{}
% ALF
\caption{<$Q_{zz}$> for Au+Au at 60 A.MeV in the \et bimodality bin,
calculated for the second largest charges $Z_{max2}$ of all events
taken all together (around their average momentum),
as a function of $a_2$.
We compare INDRA data (symbols) with filtered (dashed lines)
and unfiltered (full lines) QMD predictions for Au+Au collisions at 60 A.MeV.
 } \label{gzza2}
%%%%
\end{figure}
A thermal system has to be isotropic in its rest system. The system
we study here is the ensemble of all QP particles, i.e. those which are emitted at angles
$\theta_{cm}<90^{o}$. The degree of isotropy can be studied with help
of the momentum tensor in the rest system of the source \be
Q_{zz}=\frac{2<p_z^2>-<p_y^2>-<p_x^2>}{<p^2>}.\label{qzz}\ee $p_z$
is the momentum in the beam direction. $Q_{zz}=0$ if in the rest
system of the source the distribution is isotropic. Negative values
indicate a preferred emission in transverse direction.
Fig.\ref{figqzz} gives an overview over  $Q_{zz}$ of light charged
particles (Z=1,2) and fragments ($Z\ge 3$) in the bimodality \et
interval.

The average longitudinal velocity of all quasi projectile nucleons
in the c.m. depends on $a_2$. In the bimodality scenario this should not
be the case. It is therefore not meaningful to analyze
all events of the bimodality \et interval in a common rest frame.
We concentrate here on the question whether for a given $a_2$ the events are
isotropic. This a necessary but - as just mentioned -
not a sufficient condition for bimodality.
To obtain the $Q_{zz}$ distribution, we adopt the following
procedure:  We select events which fall into the \et
interval which shows bimodality and calculate for each $a_2$ bin the average velocity of
a) all LCP (Z=1,2) and b) of all fragments $Z \geq 3$
of all events which fall in this bin. In
one of the rest systems we calculate subsequently for each event
$Q_{zz}$ of either the fragments or the LCP.
This procedure is applied to eliminate the
dependence of the velocity of the rest system of the fragments on
$a_2$ (see fig. \ref{exZmax}), because we expect isotropy only in the rest system.
>From top to bottom, we display the $Q_{zz}$ distribution of the LCP
in the rest system of the LCP, that of fragments in
the rest system of the fragments and that of fragments in
the rest system of the LCP.  From left to right we show the unfiltered, the filtered
QMD predictions, and the INDRA data, respectively. Please note that
the \et interval is determined by the energy of the Z=1,2 particles
in the QT domain, whereas here we study the properties of Z=1,2
particles in the QP domain. Thus, autocorrelations between \et and
the particles studied in the QP domain are minimized.

We see from the top figures that the LCP in this
\et interval are preferable emitted into transverse direction in
their rest frame ($Q_{zz} < 0 $). The anisotropy depends only
slightly on $a_2$. It may be due to the fact that not only
the bimodal system but also preequilibrium emission
contributes to the spectra. It indicates that the origin of the Z=1,2 is
never a pure thermal source,  neither
for small $a_2$ nor for large $a_2$ events. The experimental
filter changes little as far as the LCP anisotropy is concerned and the
filtered QMD predictions agree quite well with the INDRA data.
In their proper rest system the fragments are emitted preferably in
forward/backward direction. The experimental filter brings the QMD predictions
of the average $Q_{zz}$ value closer to zero, in agrement with the
experiment. At very large
values of $a_2$, the emission becomes isotropic. Such a isotropy is
expected for example for the emission of a light charged fragment
from a compound nucleus. For small
$a_2$ values, $Q_{zz}$ of the fragments in the rest system of the LCP
fluctuates around 0 as
expected for an isotropic source. Thus small $a_2$ events come closest
to isotropy. For $a_2>0.5$, $Q_{zz}$ is close to
two. This means that in the LCP rest system, the largest
fragment is preferably emitted in the beam direction. This observation rules out,
on the other hand, the
hypothesis that the ensemble of fragments and light charged particles at large $a_2$
values can be considered as a pure liquid phase. This would require
that the direction of the velocity of the largest fragment is randomly
distributed in the rest system of the liquid.

One may argue that the largest fragment is not really in thermal
equilibrium in the sense that it has not lost completely its memory
on the entrance channel, i.e. its initial velocity direction. Such an argument cannot
be put forward for the second largest fragment. Therefore, we
display in fig.\ref{gzza2} the average $Q_{zz}$ for the second
largest fragments calculated in the rest system of those fragments.
The INDRA data (symbols) as well as the unfiltered (full line) and
filtered (dashed line) QMD events show  $Q_{zz} \le 0$ independent
of $a_2$. The value of $Q_{zz} \approx -0.2$ for large $a_2$ indicates
that the emission of the second largest fragment is essentially
random but still preferably in transverse direction in the fragment
rest system. For the small $a_2$ values we find $Q_{zz} \approx -0.4$.
Again, the second largest fragment is preferably emitted in
transverse direction. For the interpretation, we have to combine this
result with that shown in the lower right panel of fig. \ref{figqzz}:
in the rest system of the fragment, the emission of {\it all}
fragments is almost isotropic. A small value of $a_2$ means that the two
biggest fragments have about the same size. Therefore, combining $Q_{zz}$ of
all fragments and of the second largest fragments yields the
following scenario: For small $a_2$ values the fragment with
the highest charge has the largest velocity in beam direction, but
this time the lighter fragment can balance the momentum and
therefore the total emission pattern {\bf appears} to be isotropic,
although, if one looks into the detail, it is not.

Thus, only for the largest $a_2$ values, the fragments are
isotropically distributed in their rest system as they should if the
QP system represents the vapor or the liquid phase of a system in thermal
equilibrium. This observation for large $a_2$ values
is also compatible with the emission of a light fragment
from a compound nucleus.

\begin{figure}
\begin{center}
\includegraphics[width=16.8cm]{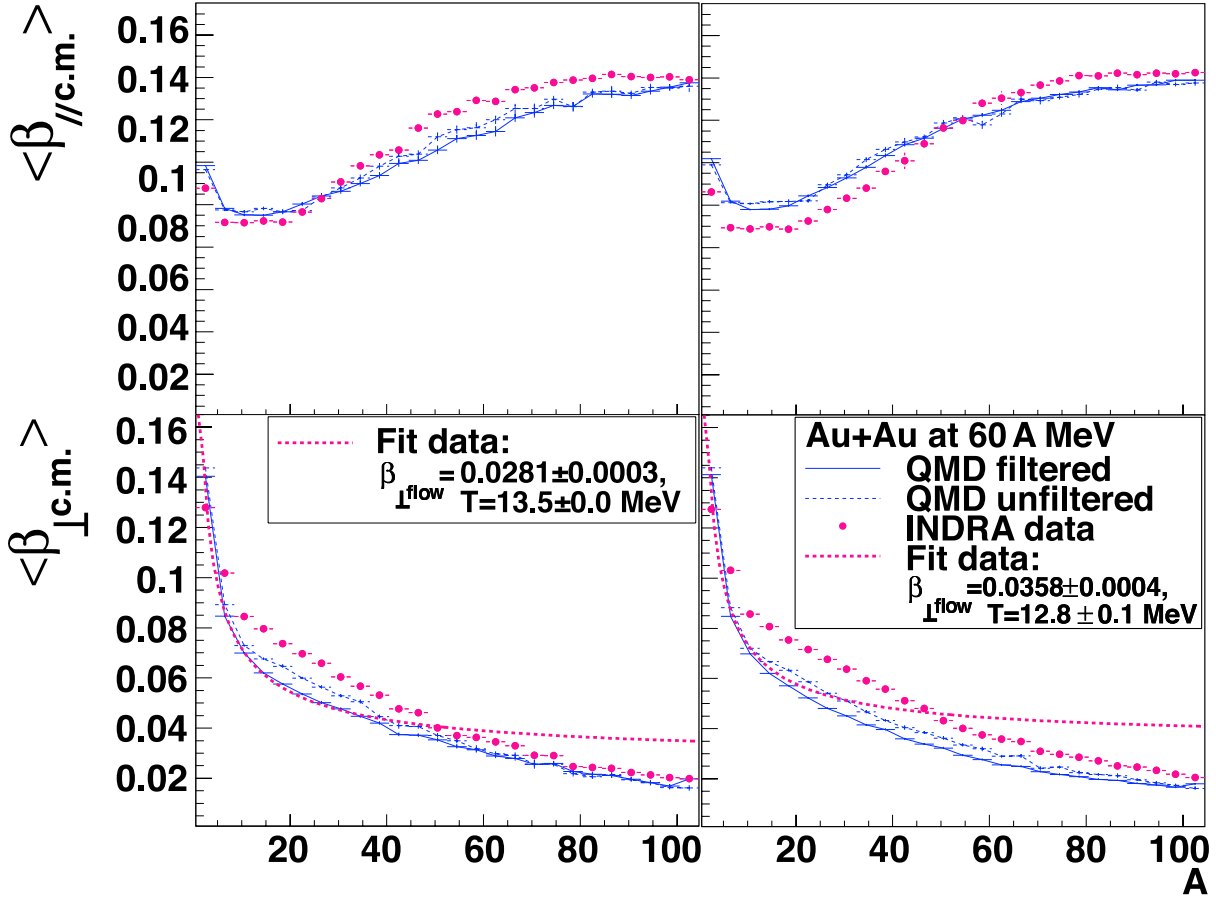}
\end{center}\emph{}
\caption{(Color online) Average longitudinal $<\beta_\parallel>$ and
average transverse velocity $<\beta_\perp>$ in the reaction center
of mass system as a function of the fragment mass. In the left panels,
we display these quantities for those
fragments which belongs to events in the bimodality
interval of \et , and in the right panels for those
of all events without any \et cut.
INDRA data (symbols) are compared with filtered (dashed lines) and unfiltered
(solid lines) QMD predictions for Au+Au collisions at 60 A.MeV. The
dashed bold line is the result of the fit of the experimental data,
using the approach of ref. \cite{hart}} \label{exA}
\end{figure}

Fig. \ref{exA} shows the average velocity in longitudinal
$\beta_{\parallel \ cm}$ and transverse direction $\beta_\perp$ in
the reaction c.m. system as a function of
the fragment mass, for those events which fall in the bimodality
interval of \et (left), and for all events (right).
In all figures we display the INDRA data as points
and compare them to the unfiltered (full
line) and filtered (dashed line) QMD predictions. Similar to fig.
\ref{exZmax}, we observe also here an increase of $\beta_
\parallel$ with increasing fragment mass. For large fragments the velocity
approaches that of the beam ($\beta_\parallel=.179$). The fragments in the
selected \et bin show - as $ Z_{max1}$ - an almost identical
behavior to those observed without a selection in \et ,
and follow the Morrissey systematics
\cite{mor}. QMD simulations reproduce $\beta_\parallel$ rather well.
The transverse velocity of the heavy fragments is small and
increases with decreasing fragment mass. Such a dependence has been
observed for systems with a collective radial velocity \cite{hart}.
In such a scenario, the form of the velocity dependence of the mass allows for the
determination of the temperature and of the radial velocity  if the
emitting source is thermal with a collective radial velocity
\cite{hart}. In a thermal system,  $E_\perp = kT$. The dotted curve,
calculated according to the formula given in ref.\cite{hart}, gives best
agreement with  the experimental data for  $E_\perp=13.5 MeV$.
This value is too high to be compatible
with the assumption of a thermalized system. At such a temperature,
the fragments would not exist anymore.

On the contrary, the observed value of 13.5 MeV is well described by
the Goldhaber model \cite{gold,mor}. It is based on the assumption
that nucleons or fragments are removed rapidly from a {\it cold}
nucleus and therefore their momentum distribution is given by the
Fermi motion. Then, the average transverse squared momentum per
nucleon of a fragment of size $A_F$ is given by the Goldhaber
formula \cite{gold}: \be p^2_\perp (A_F) \approx
\frac{2}{3}\cdot\frac{3}{5}p_F^2A_F\frac{A_P-A_F}{A_P-1} \label{gol}
\ee where $A_P$($A_P$) is the projectile (fragment) mass
and $p_F$ denotes the Fermi
momentum. Therefore \be E_\perp(A_F) = \frac{p^2_\perp
(A_F)}{2A_Fm_N}\approx \frac{1}{5}\frac{p_F^2}{m_N} \approx 14 MeV
\ee \noindent is almost independent of the fragment size and in
agreement with the data. The QMD simulations reproduce the form and
the absolute value of $\beta_\perp(A) $. This is not astonishing,
because in this model, fragments are surviving initial correlations
which preserve approximately the transverse momentum they had
initially, and therefore, the average transverse energies of
intermediate mass fragments are, independent of the impact
parameter, close to the value expected from the Goldhaber formula.
Thus, the fragment average transverse velocities are understandable
if one assumes that there is a collective radial expansion of the
system which is superimposed to an average transverse energy, given
by the Goldhaber model.

\begin{figure}
\begin{center}
\includegraphics[width=8.6cm]{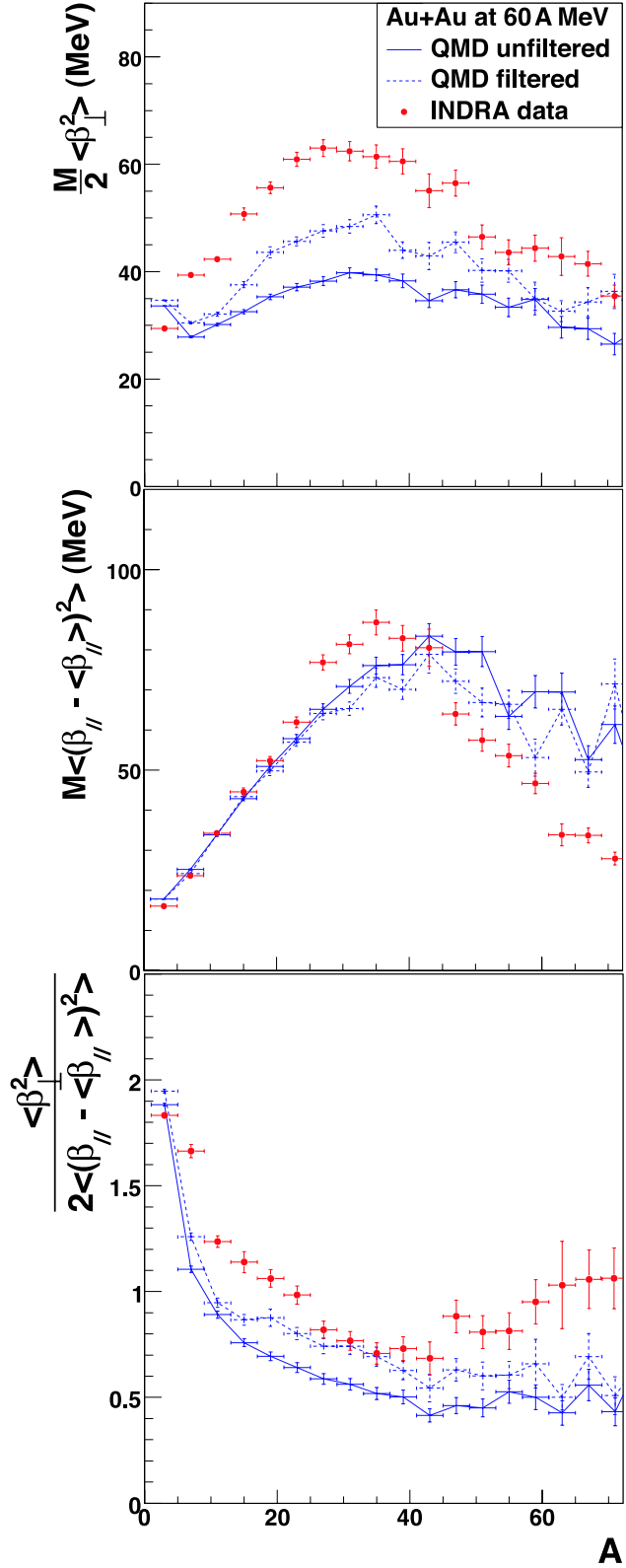}
\end{center}\emph{}
\caption{(Color online) Twice the average longitudinal (top right)
and the transverse (top left) kinetic energy of QP products in the
QP center of mass system, as well as the ratio of both (bottom), as
a function of their mass. We compare INDRA data with filtered
(dashed lines) and unfiltered (full lines) QMD predictions for Au+Au
collisions at 60 A.MeV in the bimodality \et region.} \label{etrans}
\end{figure}

Fig. \ref{etrans} displays the experimental average transverse
energy, $E_\perp$, as a function of the fragment mass of QP
products. We see indeed that $E_\perp$ in unfiltered QMD predictions
is almost independent of the fragment size. Filtering for the
experimental acceptance increases $E_\perp$ for intermediate mass
fragments and brings the calculation closer to the experimental
observation. $E_\parallel$, on the contrary, is strongly mass
dependent in the rest system of the QP. Up to A=50, this dependence
is well reproduced by QMD predictions. Above, there are
discrepancies.

A thermal system has to be isotropic in coordinate and momentum
space, and for each degree of freedom, the average kinetic energy of
the fragments has to be $E = \frac{1}{2} kT$ in the rest system of
the source. Therefore it is meaningful to calculate the deviation of
\be R = \frac{<\beta_\perp^2>}{2<(\beta_\parallel -
<\beta_\parallel>)^2>}\ee from one. The INDRA data as well as the
QMD predictions, displayed in fig. \ref{etrans}, show that R strongly depends
 on the fragment size. Up to mass A=40 , R decreases
strongly and increases slightly above. Therefore, this specific
bimodality \et interval shows the same behavior which has been
observed in \cite{Andronic:2006ra} for central collisions, where
bimodality does not occur. The discrepancy above A=40 between the
experimental data
and the QMD predictions disappears if one rejects the symmetric and
asymmetric fission events (by requiring that the product of the two
largest charges is smaller than 600). Thus, these deviations come
from events in which two fragments of similar charge are observed.

\section{System size dependence of bimodality}
\begin{figure}[ht]
\begin{center}
\includegraphics[width=8.cm]{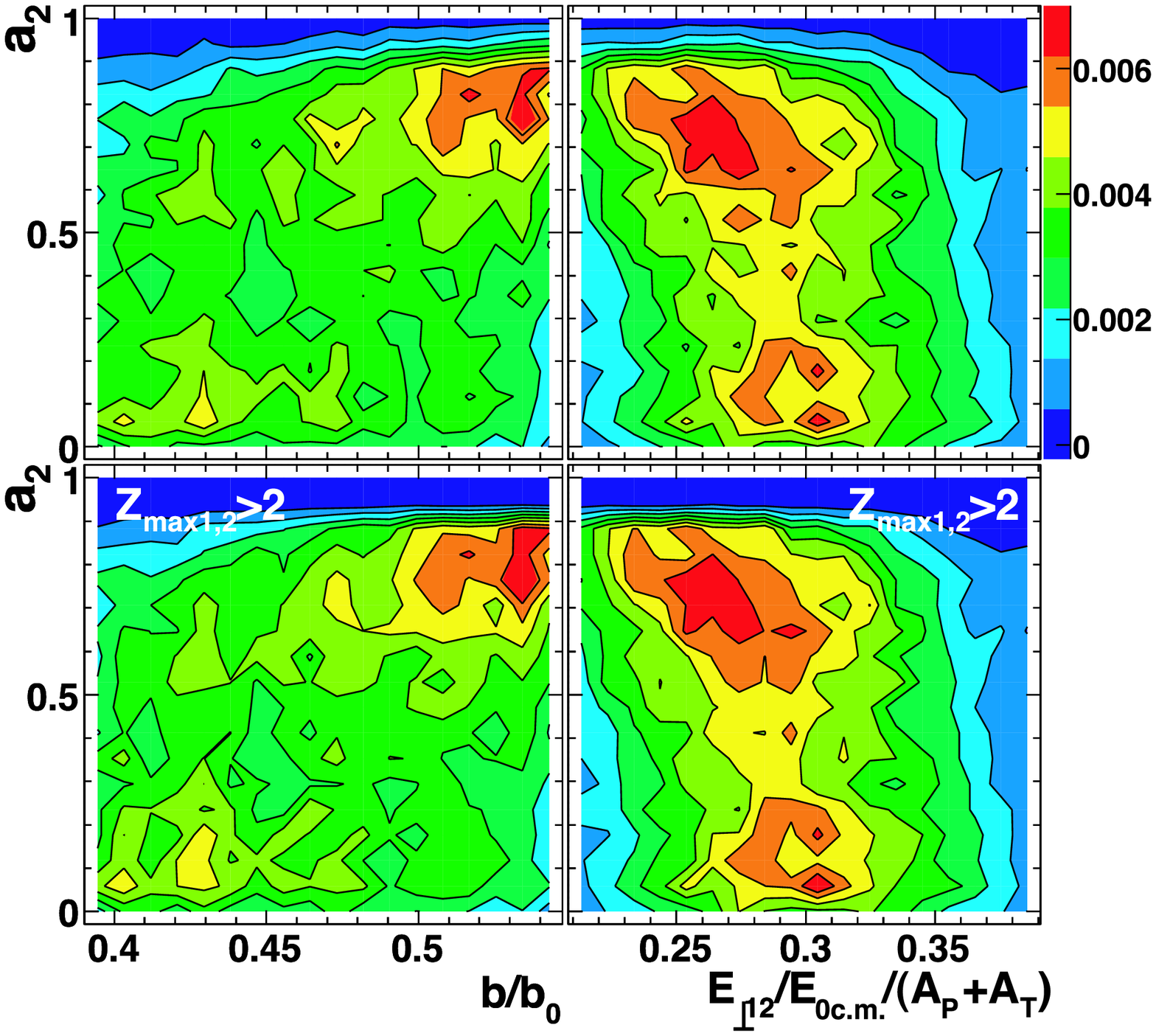}
\includegraphics[width=8.cm]{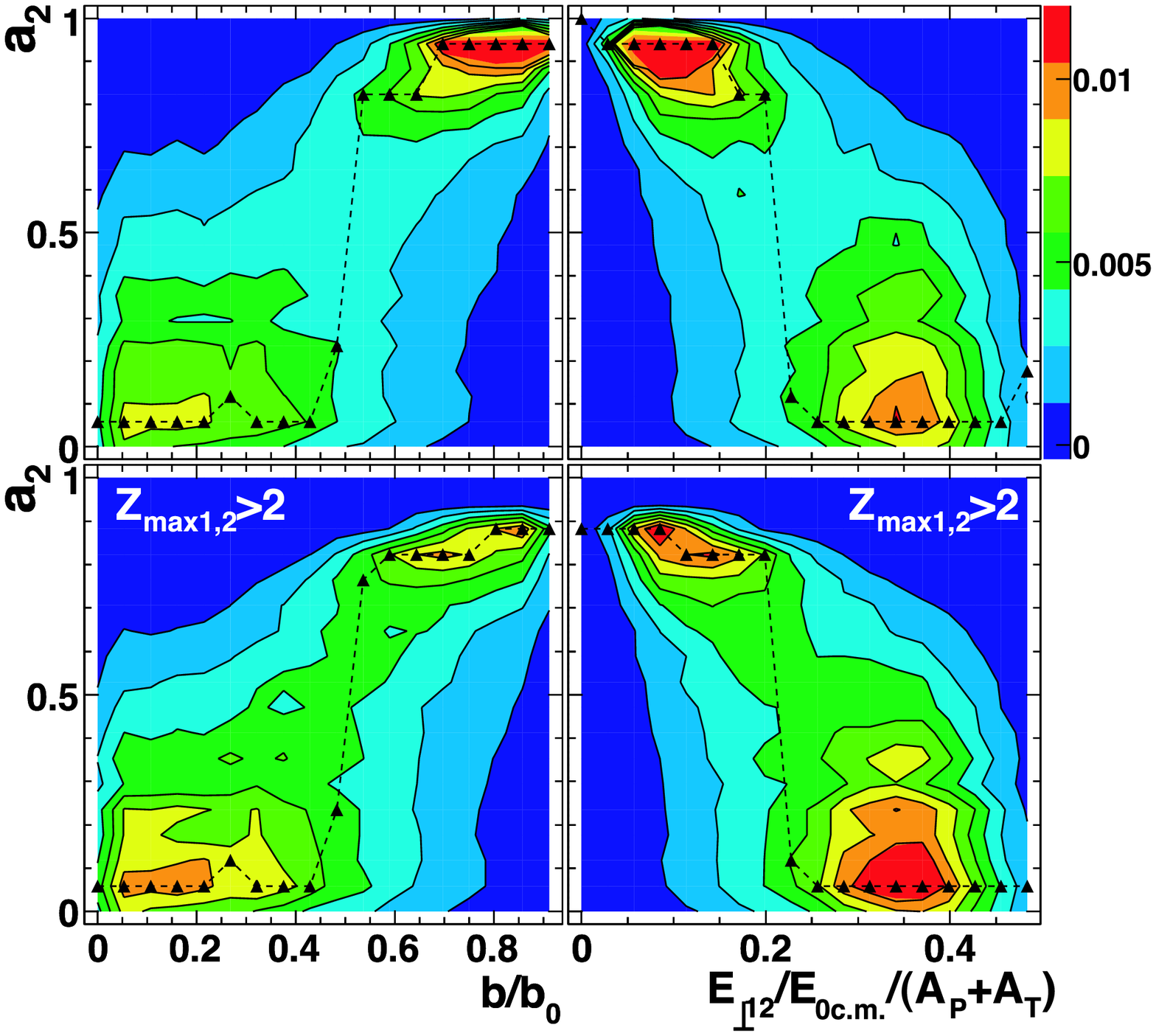}
%\end{center}\emph{}
%\caption{(Color online) The number of events as a function of $a_2$
%and of the impact parameter $b/b_0$, left, and of \etall , right. The
%number of events is given by the color coding. On the top row we
%accepts all events, on the bottom row only those with  $Z_{max} >2$
%and $Z_{max-1}>2$. The left figure shows 100 A.MeV Au+Au reactions,
%the right figure Xe+Sn reactions at the same energy.} \label{exs1}
%\end{figure}
%\begin{figure}[ht]
%\begin{center}
\includegraphics[width=8.cm]{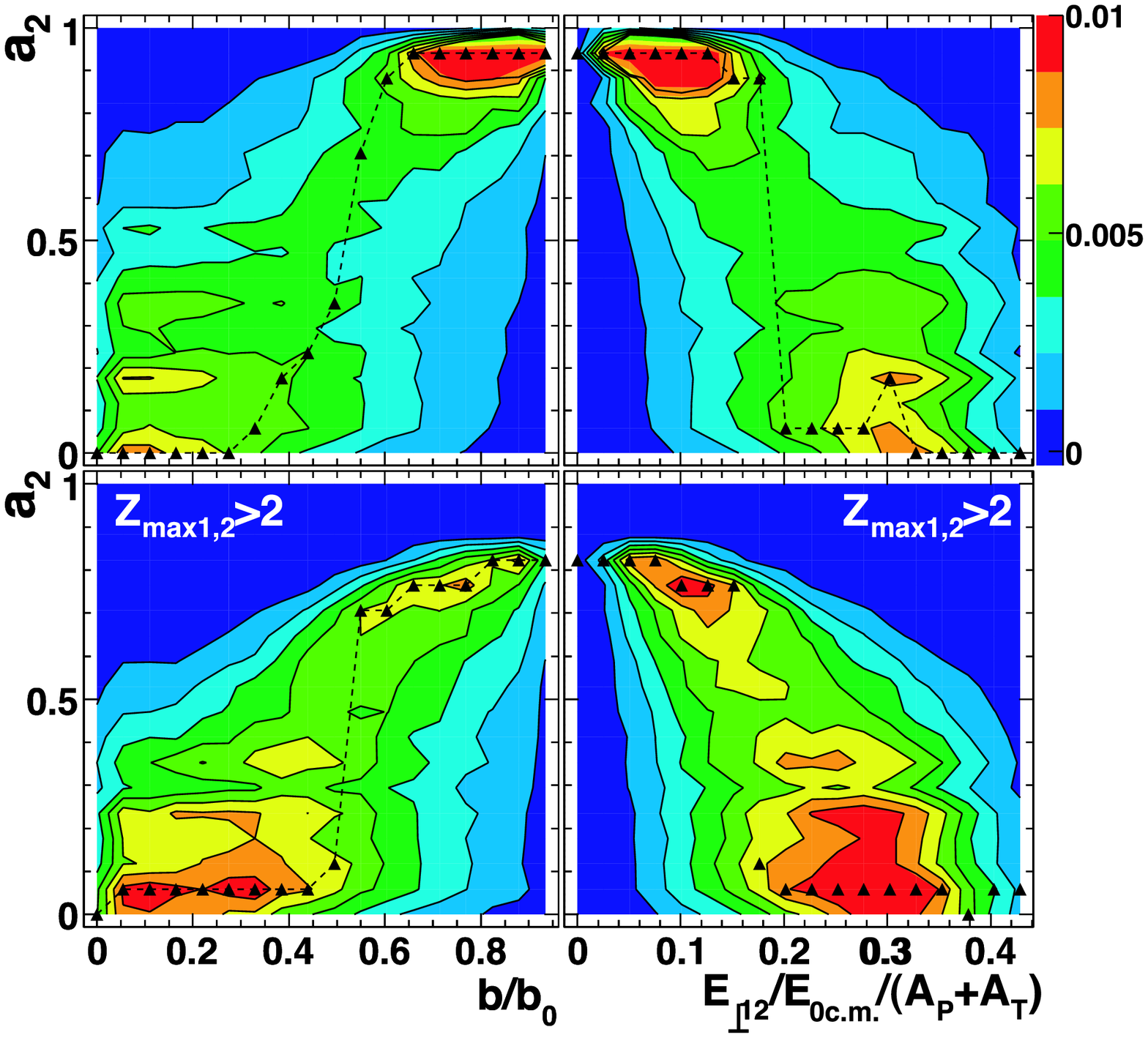}
\includegraphics[width=8.cm]{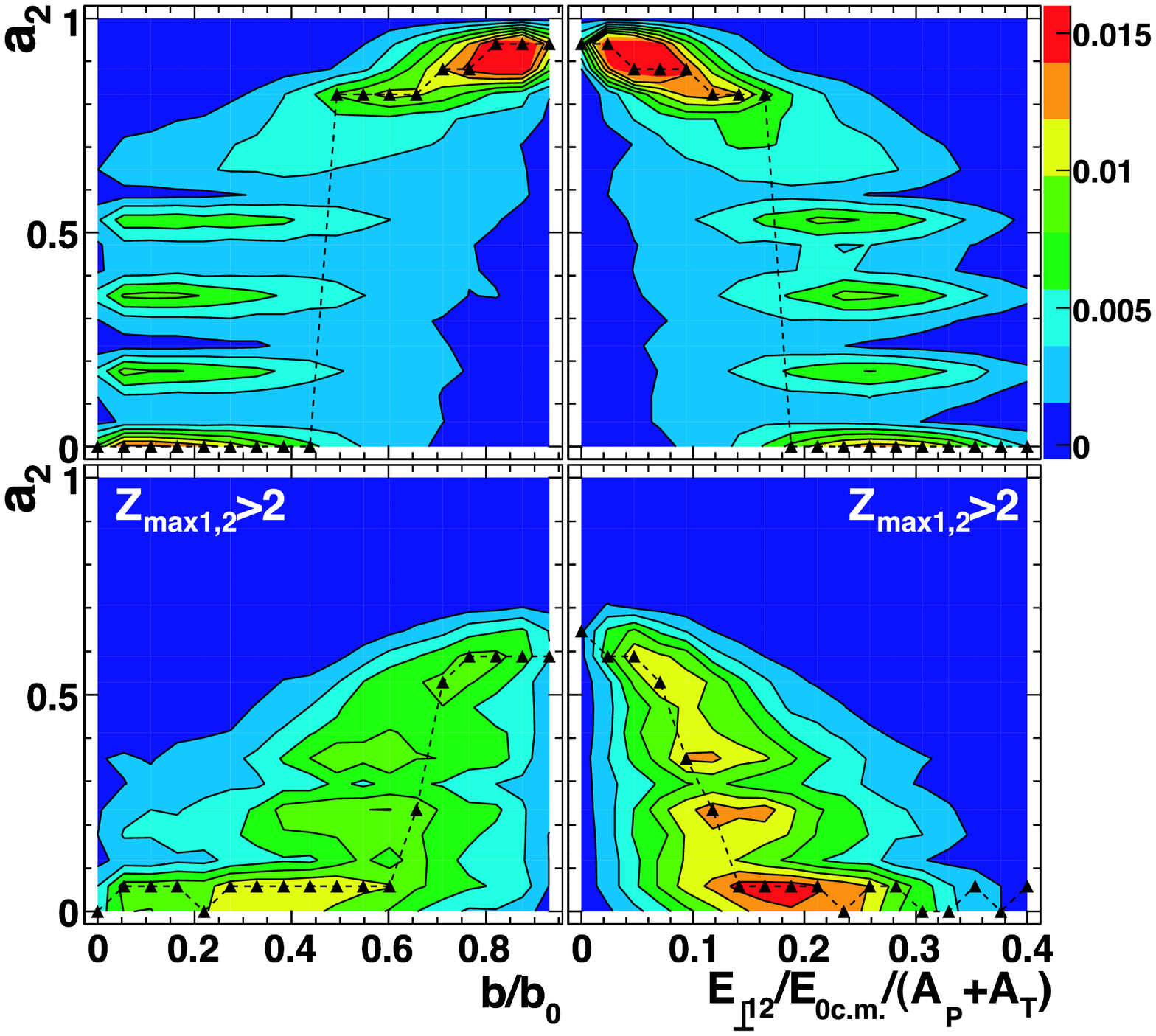}\end{center}\emph{}
\caption{(Color online) Double differential reaction cross section
(linear color scale, normalized to the number of events) as a
function of $a_2$ and of the reduced impact parameter $b/b_0$ and
the reduced \etall (see text), respectively. The symbols show the most
probable value of $a_2$. The dotted line is to guide the eye. In the
top row of each panel, all events are accepted, in the bottom row
only those with $Z_{max1} >2$ and $Z_{max2}>2$ are shown. We display
Au+Au (top left),
Xe+Sn (top right), Kr+Kr (bottom left) and
Ca+Ca (bottom right) reactions at 100 A.MeV incident energy.
For Au+Au, note that a narrow selection in impact parameter ($5 fm \leq b
\leq 7 fm$)
has been applied around the bimodality region.} \label{exs2}
\end{figure}

After having established that most of the data for Au+Au collisions
at 60 A.MeV are quantitatively described in QMD calculations, we
study now the system size dependence of the bimodality.
Unfortunately, no data have been published so far to verify these theoretical
predictions. In order to discuss the physics, we present, in fig.
\ref{exs2}, $a_2$ as a function of the reduced \etall (right) as well as
a function of the reduced impact parameter $b/b_0$ (left). Here, \etall
is calculated over all particles with charge Z=1,2 (QT and QP) and
$a_2$ is given for QP products. To account for system energy and size
scalings, \etall is divided by the total mass of
the system and by the energy per nucleon in the center of mass. The
top row shows $a_2$ calculated with all fragments. For the bottom
row, we require that the fragment with the largest charge,
$Z_{max1}$, and that with the second largest charge $Z_{max2}$ have
both a charge larger than 2.
\begin{figure}[ht]
\begin{center}
\includegraphics[width=8.6cm]{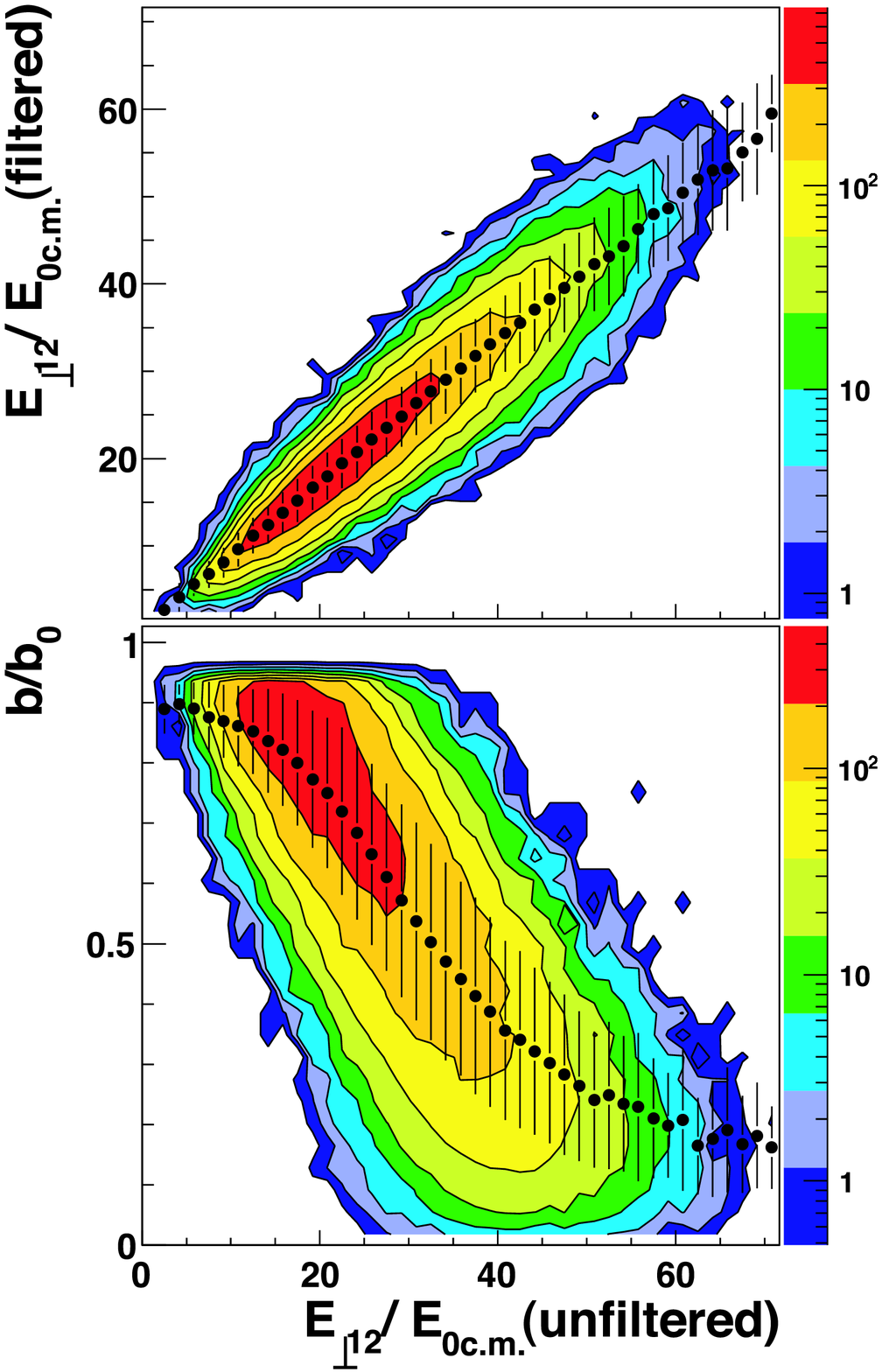}\end{center}\emph{}
\caption{(Color online) Double differential reaction cross section
(logarithmic color scale in arbitrary units), for Au+Au at 60 A.MeV
QMD predictions, as
a function of  \etall divided by the kinetic energy per nucleon in the
center of mass of the reaction system for INDRA filtered events (top) and
as a function of the reduced impact parameter $b/b_0$ (bottom),
respectively,  and as a function of the unfiltered reduced \etall.
The symbols show the mean values of
\etall $/E_{0cm}$ and the error bars the rms of their distribution.}
\label{flu}
\end{figure}
We start the discussion with the Au+Au reaction (top left) for which
we zoom on the narrow impact parameter interval $5\ fm \le b\le 7\
fm$, where bimodality occurs. If plotted as a function of \etall for
this impact parameter interval, we
clearly see the two event classes with a distinct $a_2$ value for
the same small \etall interval. Whether we limit the events to $Z_{max1}>2$
and $Z_{max2}>2$ or not does not make a difference, because there
are almost always two fragments with $Z\ge 3$. When plotted as a
function of $b/b_0$, the bimodality structure with a sudden jump
has disappeared. There, the
events with a small $a_2$ value are distributed over a broad range
of impact parameters \cite{zbiri}. Because the fluctuations in \etall
for a given impact parameter are large, some of these events appear
in the same bimodality interval in \etall as the events with a large
$a_2$. This is shown in fig.\ref{flu} which displays the filtered INDRA
\etall distribution and that of $b/b_0$ for a given unfiltered
\etall in QMD simulations of Au+Au reactions at 60 A.MeV. For an
incident energy of 100 A.MeV, the distributions are similar. The bars
mark the standard deviation. We see that there is a strong
correlation between these observables. In particular, the INDRA
set-up provides a mean linear response - no saturation - to \etall over
a large range of \etall (hence to the multiplicity and to the energy
of particles).
The distributions are, however, quite broad, and hence, for a given
experimentally measured \etall value, the unfiltered \etall 's as well as
the impact parameters show large fluctuations.

Like Au+Au, the smaller Xe+Sn system (fig.\ref{exs2}, top right)
exhibits two distinct maxima of $a_2$,
with a sudden jump of the most probable value of $a_2$
 (depicted by the dashed lines).
The two event classes are also seen if the events are plotted as a
function of $b/b_0$, and it is visible that they are both associated
with quite different impact parameters. Thus, nuclei disintegrate in
two quite distinct pattern, but they belong to quite different
impact parameters, and hence to quite different reaction geometries. For the even
smaller Kr+Kr system (fig.\ref{exs2}, bottom left), there is still a
sudden jump of $a_2$ as a function of \etall, but the two event classes
become less distinct, the relative yield of intermediate $a_2$
values getting higher. This can be explained by the fact that the
absolute value of \etall is reduced and hence the relative fluctuations
around the mean value increase for a given $b/b_0$. Therefore, we
find again the existence of the two maxima in $a_2$ for (almost) the
same value of the reduced \etall if $Z_{max1}
>2$ and $Z_{max2}>2$. The impact parameter of the two event classes
is, however, quite different, the minimum of $a_2$ is less pronounced.
In addition, events with $Z_{max1}>2$ and $Z_{max2}>2$ become rare.
The majority of events with a large $a_2$ value are now those in
which the second largest fragment charge has $Z=2 $. For such small
systems, already a beam energy of 100 A.MeV makes the reaction that
violent that in central collisions fragments do hardly survive.
Finally, for the very small system Ca+Ca (fig.\ref{exs2}, bottom
right), bimodality becomes almost impossible, many intermediate
values of $a_2$ are highly populated, because the system is too
small. For the rare events with $Z_{max1}>2$ and $Z_{max2}>2$, large
values of $a_2$ are impossible, and therefore we cannot have two
distinct maxima anymore. The reaction is dominated by $\alpha$
emission from the QP, as can be seen in the top row.

It is
remarkable that, independent of the system size, the sudden jump of
the mean $a_2$ value occurs around \etall $/ E_{0cm}/ (A_P+A_T) \approx
0.2$, i.e. when the transverse energy of light charged particles per
nucleon is identical. This scaling is understandable, because this
quantity measures the energy transfer in the reaction and extends
the scaling we have observed already for the beam energy dependence
of the bimodality for the Au+Au system \cite{tam,arn}.

It is an important observation that nature disfavors intermediate
$a_2$ ($a_2 \approx 0.5$) values. Either a big cluster emits small
fragments or we observe multifragmentation, events in which several
intermediate mass fragments are produced. The dominance of these two
reaction scenarios is independent of the system size. Dynamical
models reproduce this observation but it is very desirable to know
whether also statistical models predict such a suppression of
intermediate $a_2$ events. This would allow to elucidate whether
this suppression is related to phase space or to the nucleon-nucleon
interaction.

Independent of the system size, the events with large $a_2$ belong
to other impact parameters than those with a small $a_2$. But \etall
and the impact parameter are not well correlated and therefore, due
to the fluctuations of \etall for a given impact parameter, events with
very different $a_2$ values appear in the same \etall interval. For
smaller systems, the events with large $a_2$ are those in which an
$\alpha$ particle is emitted from the residue. There we see as well
that events with different $a_2$ do not belong to the same \etall
interval.

\section{Energy dependence of bimodality}
\begin{figure}[ht]
\begin{center}
\includegraphics[width=8.6cm]{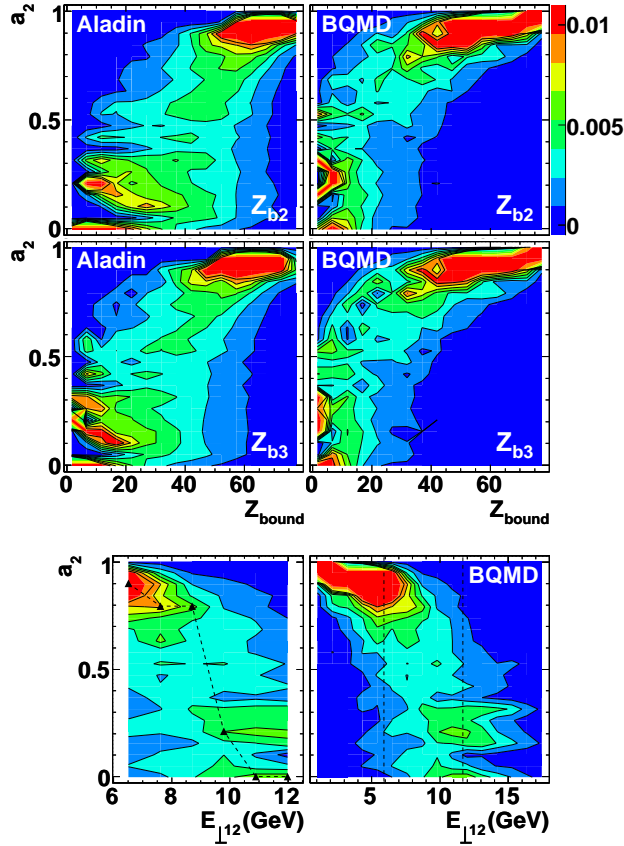}
\end{center}\emph{}
\caption{(Color online) $a_2$ of the projectile spectator
as a function of $Z_{b2}$ (top) and of
$Z_{b3}$ (middle) for Au+Au collisions at 600 A.MeV  \cite{alad}. On
the left hand side, we show the Aladin experimental results, on the right
hand side the QMD predictions. The bottom row displays the
theoretical predictions of $a_2$ as a function of \etall (right) which
are enlarged around intermediate \etall values (delimited by dashed
lines) in the left panel. In this latter panel, the symbols show the
most probable value of $a_2$, and the dotted line is to guide the eye.
}\label{ala}
\end{figure}
If bimodality is a special manifestation of the general feature
that, in heavy ion reactions, two distinct reaction scenarios with
quite different $a_2$ values exist, it is tempting to see whether
this observation continues to higher energies. Whereas at low
energies multifragmentation happens in central collisions, and
therefore fragments are formed from the participating nucleons,
at higher energies \cite{zbiri},
multifragmentation happens at large impact parameters and fragments
are formed from spectator nucleons. There are not many experimental results
available which allow to study this question, especially there is no
experiment in which simultaneously \etall and all fragment charges have
been measured in the target or projectile spectator.
The only experiments which allow to address this
question is the ALADIN experiment at GSI. In this experiment, Au +
Au at 600 A.MeV incident energy \cite{alad}, almost all fragments with $ Z\ge 2$ of the
projectile spectator (PS), i.e. in forward direction, have been
measured, but no light charged particles. Because $Z=1$ particles are
not measured, \etps cannot be extracted, and we have to conclude the
existence of bimodality indirectly. In ref. \cite{kre}, it has been
shown, however, that the inclusive (impact parameter averaged)
yield of $a_2$ has maxima at small and large values, separated
by a minimum at around
$a_2=0.5$, similar to the observations at energies around 100 A.MeV.

These data have also been analyzed by a statistical model approach
\cite{bot} where it has been shown that the experimental mean values
and fluctuations are well described once the distribution of the
system energy, E, has been adapted. It has, however, not been
demonstrated that this energy distribution corresponds to that which is
expected for a given temperature of the system, as required by the
bimodality assumption. In any case, the experimental inverse slope
parameters are much larger than those of statistical model
calculations \cite{odeh}. These experimental
parameters are compatible with the prediction of
the Goldhaber model (eq. \ref{gol}).

In a first step, we have to verify that QMD simulations reproduce
correctly the pattern of $a_2$ as  a function of $Z_{bound}$, the
measured charge of all fragments with $2\le Z\le 30$ ($Z_{b2}$) or $3\le
Z\le 30$ ($Z_{b3}$) in the PS. In the ALADIN experiments, it has
been shown that $Z_{bound}$ is strongly correlated with the energy
deposit during the reaction \cite{alad}. In a second step, we
replace then $Z_{bound}$ by \etall (calculated with both target and
projectile spectator particles). Because $Z_{bound}$ as well as \etall
are considered as a good mesure for the centrality of the
reaction, and more precisely of the energy deposit in the spectator,
such a replacement is meaningful. Fig. \ref{ala} shows in the two
top rows $a_2$ as a function of the bound charge in the PS domain.
On the left hand side, we display the results obtained with the
ALADIN set-up, on the right hand side the filtered QMD events.
Filtering is here not a very important issue, because the ALADIN set-up
registers the large majority of the fragments in the PS region.
$a_2$ as a function of $Z_{b2}$ and $Z_{b3}$ is shown in the top and
middle panel, respectively. We see a quite reasonable agreement
between theory and experiment. This allows us to replace in the
bottom row $Z_{bound}$ by \etall. On the right hand side of the bottom
row, we see that also at 600 A.MeV bimodality can be observed. In a
\etall interval around 9 GeV, we see two event classes: one with a
large and one with a small value of $a_2$. The left hand side of
the bottom row zooms in this \etall interval. As at beam energies
around 100 A.MeV \cite{zbiri}, the two event classes are separated
by a region with $a_2 \approx 0.5$ which contains only a very
limited number of events.

\section{conclusion}
The appearance of two distinct reaction scenarios,
multifragmentation and residue production for the {\it same} value
of \etall is a very genuine phenomenon in heavy ion collisions. It
exists over almost the whole energy range for which
multifragmentation has been observed, and it exists in participant
fragmentation as well as in spectator fragmentation. The fact that
physical events show either a small or a large $a_2$ value but
almost never an intermediate $a_2$ value is first of all remarkable.
It is also very general, and the classification of events with the help
of $a_2$ is a good way to elucidate this fact. Whether phase space
or nuclear interactions are at the origin of the lack of events with
intermediate $a_2$ values is still unknown.

In the dynamical QMD model, the large impact parameter fluctuations
for a given \etall are the reason that events with small and large
$a_2$ values appear for the same value of \etall . The data are in
agreement with predictions of models which assume that
multifragmentation is a fast process. QMD simulations, in which
fragments are surviving initial correlations, reproduce the large
majority of the experimental observations.

The investigation of the bimodality \etall interval shows that the
majority of the events in this interval has properties which are not
compatible with the assumptions that large and small $a_2$ events
belong to two phases which exist simultaneously in a small interval
of the temperature measured by \etall. We can, however, not exclude
that a subset of the events shows the properties expected from a
statistically equilibrated system. Investigations whether such a
subset can be found have been advanced recently \cite{indr}. It would
therefore be interesting to see whether such a subset may be a sign
of true bimodality, i.e. can be reproduced in statistical models
with an energy fluctuation expected for a system having a fixed
temperature.
Of course generalized statistical ensembles can be defined \cite{gor}
assuming that in each event E,N,V differ. It remains to be seen 
how the distribution of E,N,V can be assessed and whether 
such an approach is compatible with data.

\end{document}